\documentclass[prl,showpacs,twocolumn]{revtex4}
\usepackage{graphicx}

\begin{document}

\title{Ferroelectricity in spiral magnets}
\author{Maxim Mostovoy}
\affiliation {Materials Science Center, University of Groningen,
Nijenborgh 4, 9747 AG Groningen, The Netherlands\\and\\ II
Physikalisches Institut, Universit\"at zu K\"oln,
Z\"ulplicherstrasse 77, D-50937 K\"oln, Germany}

\date{October 26, 2005}


%77.80.-e    Ferroelectricity and antiferroelectricity
%75.30.Fv    Spin-density waves
%75.60.Ch    Domain walls and domain structure
%77.80.Fm    Switching phenomena

\begin{abstract}
It was recently observed that materials showing most striking
multiferroic phenomena are frustrated spin-density-wave magnets.
We present a simple phenomenological theory, which describes the
orientation of the induced electric polarization for various
incommensurate magnetic states, its dependence on temperature and
magnetic field, and anomalies of dielectric susceptibility at
magnetic transitions. We show that electric polarization can be
induced at domain walls and that magnetic vortices carry electric
charge.
\end{abstract}

\maketitle

The recent revival of interest in materials showing strong
interplay between magnetism and ferroelectricity led to the
discovery of a new class of systems, which can be called
ferroelectric magnets. While in `old' multiferroics, such as
BiFeO$_3$, spontaneous electric polarization appears at a much
higher temperature than magnetism, in RMnO$_3$ (R = Tb, Dy, Gd)
\cite{Kimura,Kimura2003,Goto,KimuraLawes}, RMn$_2$O$_5$ (R = Tb,
Ho, Dy)\cite{Higashiyama,Hur1} and Ni$_3$V$_2$O$_8$ \cite{Lawes}
ferroelectricity is only observed in magnetically ordered states.
In magnetic fields these materials show reversals and sudden flops
of electric polarization vector
\cite{Kimura2003,KimuraLawes,Hur1}, and an exceptionally strong
enhancement of dielectric constant (the giant magnetocapacitance
effect) \cite{Goto}. These remarkable multiferroic phenomena,
essential for control of dielectric properties by magnetism,
follow from the fact that that ferroelectricity in these materials
is induced by magnetic ordering.

An intriguing feature of the new class of multiferroics is an
intimate link between the spontaneously induced polarization and
magnetic frustration. In RMnO$_3$ a large distortion of the cubic
perovskite lattice gives rise to competing ferro- and
antiferromagnetic interactions between the Mn spins \cite{Kimura},
while in the Kagom\'e staircase material Ni$_3$V$_2$O$_8$
 frustration originates from the lattice geometry \cite{Lawes}. The
competing exchange interactions stabilize spin-density-wave (SDW)
states with a periodically varying magnetization \cite{Kimura}.

The relation between the ferroelectricity and incommensurate
magnetism is widely used as an empirical guiding principle in the
search for new multiferroic materials. In this paper we discuss a
phenomenological description of ferroelectric magnets, based on
general symmetry arguments. We formulate a simple continuum model,
which clarifies the relation between the induced electric
polarization and magnetic structure, describes anomalies of
dielectric constant at magnetic transitions and qualitatively
explains complex magnetic field behaviors found in these
materials. We extend this analysis to domain walls and vortices.
Our phenomenological approach is complimentary to the recent
discussions of microscopic mechanisms of ferroelectricity in
magnets \cite{SergienkoDagotto,Katsura}.

\textit{Induced polarization:} Incommensurate SDW states are
largely insensitive to details of crystal structure and can be
described by a continuum field theory of the Ginzburg-Landau type.
The form of the coupling of electric polarization ${\bf P}$ to
magnetization ${\bf M}$ can be found using general symmetry
arguments. The invariance upon the time reversal, $t \rightarrow
-t$, which transforms ${\bf P} \rightarrow {\bf P}$ and ${\bf M}
\rightarrow -{\bf M}$, requires the lowest-order coupling to be
quadratic in ${\bf M}$. The symmetry with respect to the spatial
inversion, ${\bf x} \rightarrow - {\bf x}$, upon which ${\bf P}
\rightarrow - {\bf P}$ and ${\bf M} \rightarrow {\bf M}$, is
respected when the coupling of a uniform polarization to an
inhomogeneous magnetization is linear in ${\bf P}$ and contains
one gradient of ${\bf M}$. Omitting vector indices the coupling
term in thermodynamic potential can be written in the form
\begin{equation}
\Phi_{em}({\bf P},{\bf M}) \propto P M \partial M.
\label{eq:coupling}
\end{equation}
The terms linear in gradient (Lifshitz invariants) are allowed in
systems with broken inversion symmetry, such as noncentrosymmetric
crystals, where they can give rise to periodic spatial modulations
of magnetization \cite{Dzyaloshinskii}. Such an incommensurate SDW
state is observed in the ferroelectric BiFeO$_3$, where the
inversion symmetry is spontaneously broken by electric
polarization. The helix with the long period $620${\AA} in
BiFeO$_3$ results from the coupling between the electric
polarization ${\bf P}$ and the G-type AFM order, which has the
form Eq.(\ref{eq:coupling}) with the uniform magnetization ${\bf
M}$ replaced by the slowly varying N\'eel vector ${\bf L}$
\cite{Zvezdin}. The small value of the wave vector is a
consequence of the relativistic nature of the coupling
\cite{Dzyaloshinskii,Morya}.

This reasoning can be turned around to explain electric
polarization in frustrated magnets, in which an incommensurate
magnetic ordering results from competing exchange interactions and
the SDW wave vector is, in general, not small. When the SDW order
of a proper kind sets in, the coupling Eq.(\ref{eq:coupling})
induces  a uniform electric polarization that breaks the inversion
symmetry. The weakness of the coupling translates in this case to
relatively low values of the induced polarization.

This mechanism does not require a special kind of crystal lattice.
In the simplest case of cubic symmetry the coupling term has the
form
\begin{equation}
\Phi_{em}({\bf P},{\bf M}) = \gamma {\bf P} \cdot \left[ {\bf M}
\left(\nabla \cdot {\bf M}\right) -\left( {\bf M} \cdot \nabla
\right) {\bf M}+ \ldots \right].
%\label{eq:Phime2}
\end{equation}
The omitted terms can be written as the the total derivative,
$\nabla f({\bf M})$, and do not contribute to the uniform
polarization. Assuming that in absence of magnetism the system
shows no instability towards ferroelectricity, we only keep the
quadratic term in the `electric part' of the thermodynamic
potential, $ \Phi_e({\bf P}) =\frac{P^2}{2\chi_e}$, where $\chi_e$
is the dielectric susceptibility in absence of magnetism. The
variation of $\Phi_e + \Phi_{em}$ with respect to ${\bf P}$ then
gives
\begin{equation}
{\bf P} = \gamma \chi_e  \left[ \left( {\bf M} \cdot \nabla
\right) {\bf M} - {\bf M} \left(\nabla \cdot {\bf M}\right)
%+ \nabla f({\bf M})
\right]. \label{eq:P}
\end{equation}

\begin{figure}[htbp]
\centering
\includegraphics[width=5cm]{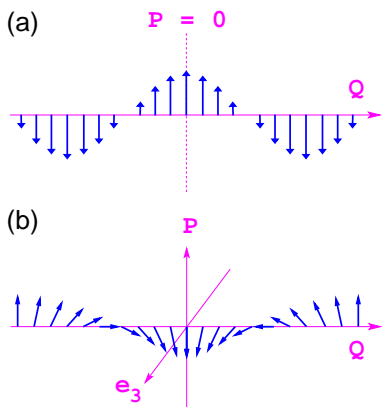}
\caption{The sinusoidal SDW [panel (a)] does not induce a uniform
electric polarization, while for a helicoidal SDW [panel (b)] the
electric polarization vector ${\bf P}$ lies in the spin rotation
plane (with the normal vector ${\bf e}_3$) and is transverse to
the wave vector ${\bf Q}$ of the helix. \label{fig:sdw}}
\end{figure}

Consider now a SDW state with the wave vector ${\bf Q}$,
\begin{equation}
{\bf M} =  M_{1} {\bf e}_1 \cos \phi +
M_2 {\bf e}_2 \sin \phi +M_{3} {\bf e}_3,
\label{eq:helix}
\end{equation}
where $\phi =  {\bf Q} \cdot {\bf x}$ and  the unit vectors ${\bf
e}_i$, $i =1,2,3$ form an orthogonal basis. If only $M_1$ or $M_2$
is nonzero, this state is the sinusoidal wave (see
Fig.~\ref{fig:sdw}a), while for $M_{1} ,M_{2} \neq 0$ it is an
(elliptical) helix with the spin rotation axis ${\bf e}_3$. If
also $M_{3} \neq 0$, then the helix is conical. Using
Eq.(\ref{eq:P}), we find that the average polarization is
transverse both to the spin rotation axis ${{\bf e}_3}$ and the
wave vector ${\bf Q}$ of the helix (see Fig.~\ref{fig:sdw}b), and
is independent of $M_{3}$:
\begin{equation}
{\bar {\bf P}} = \frac{1}{V} \int\!\!d^3x {\bf P} = \gamma \chi_e
M_{1}M_{2} \left[ {\bf e}_3 \times {\bf Q} \right].
\label{eq:Phelix}
\end{equation}
For the sinusoidal SDW the induced polarization is $0$: such an
ordering does not break the inversion symmetry at the sites where
the magnetization reaches maximum or minimum (e.g., the points on
the dashed line in Fig.~\ref{fig:sdw}a) and, therefore, it cannot
induce an average electric polarization. Equation
(\ref{eq:Phelix}) also holds for the orthorombic crystal symmetry,
provided that ${{\bf e}_3}$ and ${\bf Q}$ are parallel to crystal
axes.

This explains why  the transition at $T_{S} = 41$K to the
sinusoidal SDW state in TbMnO$_3$\cite{Kimura2003,Kenzelmann} does
not give rise to ferroelectricity. The polarization is only
induced below the so-called lock-in transition at $T_{H} = 28$K,
when the sinusoidal SDW is replaced by the helix with the
propagation vector ${\bf Q}$ parallel to the $b$ axis and the Mn
spins rotating in the $bc$ plane (${\bf e}_{3}
\parallel a$) \cite{Kenzelmann}, so that according to Eq.(\ref{eq:Phelix})
the polarization is induced along the $c$ axis in agreement with
experiment. Similarly, the polarization is absent in the
high-temperature incommensurate phase of Ni$_3$V$_2$O$_8$, which
is the sinusoidal SDW state \cite{Lawes}. The helix with ${\bf Q}
\parallel a$ and the spin rotation axis ${\bf e}_{3}
\parallel c$, which appears in the low-temperature incommensurate
phase, induces polarization along the $b$ axis.

We note that the spiral ordering can be considered as a particular
case of a magnetic ordering with two noncollinear SDWs with equal
wave vectors,
\[
{\bf M} =  M_{1} {\bf e}_1 \cos \left( {\bf Q} \cdot {\bf
x} + \phi_1\right) + M_2 {\bf e}_2 \cos \left({\bf Q} \cdot {\bf
x} + \phi_2\right),
\]
such as the one recently found in RMn$_2$O$_5$
\cite{Chapon2004,Blake}. Although a single sinusoidal SDW does not
induce polarization, the interference between the two SDWs gives
\begin{equation}
{\bar {\bf P}} = \gamma \chi_e M_{1}M_{2} \sin \left(\phi_2 -
\phi_1 \right)\left[ {\bf Q} \times \left[ {\bf e}_1 \times {\bf
e}_2 \right] \right]. \label{eq:2SDW}
\end{equation}

\textit{Magnetic textures:} Electric polarization can also be
induced in nonfrustrated magnets near magnetic defects, e.g.,
domain walls or in inhomogeneous ground states stabilized by
magnetostatic interactions, e.g., vortices in nanodiscs
\cite{Cowburn}. In many cases the integrated quantities, such as
the total polarization per unit area of a domain wall, only depend
on topology of spin textures. For walls with collinear spins and
the ones with spins rotating around the axis normal to the wall,
the total polarization is 0, while the wall, in which spins rotate
around an axis parallel to the wall,  $ {\bf M} = M \left[\cos
\phi(x_1)  \, {\bf e}_1 + \sin \phi(x_1)  \, {\bf e}_2 \right]$,
has the total polarization that depends on the total rotation
angle
\begin{equation}
\int_{-\infty}^{+\infty}dx_1{\bf P} =2 \gamma \chi_e M^2 {\bf e}_2
[\phi(+\infty) - \phi(-\infty)].
\end{equation}
Using Eq.(\ref{eq:P}) one can also show that the vortex
\[
{\bf M} = M\left[\cos (n\phi + \phi_0) {\bf e}_1 + \sin
(n\phi +\phi_0) {\bf e}_2 \right],
\]
where $\phi = \arctan \frac{x_2}{x_1}$ and $\phi_0$ is an
arbitrary phase, has quantized electric charge located at the
vortex core: $q_n = n q_1$, where $q_1 = 4\pi\gamma \chi_e M^2 $
and $n$ is the winding number of the vortex. An applied electric
field will move magnetic vortices and anti-vortices in opposite
directions.

\textit{Sinusoidal-helicoidal transition:} We now turn to the
phase diagram of ferroelectric SDW magnets. Using the values of
the induced polarization ($10^2-10^3\mu \mbox{C m}^{-2}$) and
magnetic transition temperatures ($5-40$K) for these materials, we
find that the energy gain related to the induced polarization is
small compared to the magnetic energy gain. Therefore,  the
temperature and magnetic field dependence of the polarization
merely reflects the changes in magnetic ordering, which can be
described using  the Ginzburg-Landau  thermodynamic potential
\begin{equation}
\Phi_m({\bf M}) = \sum_{i=x,y,z} \frac{a_i}{2} (M_i)^2 +
\frac{b}{4} M^4 + \frac{c}{2} {\bf
M}\left(\frac{d^2}{dx^2}+Q^2\right)^2 {\bf M}.\label{eq:onelayer}
\end{equation}
In what follows we assume that $a_x < a_y < a_z$ (easy axis along the $x$
direction) and first neglect higher-order anisotropies. The last
term in Eq.(\ref{eq:onelayer}) favors a periodic SDW ordering with
the wave vector $Q$ along the $x$ axis.

A down-shift of the ferroelectric transition with respect to the
magnetic one, found in all magnetic ferroelectrics, is a
consequence of magnetic anisotropy. While for an isotropic system
the ground state is a helix with a constant $M$, an anisotropic
system first undergoes a transition to the sinusoidal SDW state
with ${\bf M}$ along the easy axis, ${\bf M} = M_x {\hat x} \cos
Qx$, at temperature $T_S: a_x(T_S) = 0$. As temperature is lowered
and the amplitude of the order parameter grows, the system
undergoes a second transition at some $T_H < T_S$ to the
elliptical helix state, ${\bf M} = M_x {\hat x} \cos Qx + M_y
{\hat y} \sin Qx$, provided that the anisotropy parameter $\Delta
= a_y - a_x$ is not too large. When the two transitions occur at
close temperatures, the higher harmonics in the SDW state are
small and the helix appears at $a_y = a_x/3$. For $a_x(T) = \alpha
(T - T_S)$ we then obtain
\begin{equation}
T_H = T_S - \frac{3\Delta}{2\alpha}. \label{eq:Tc2(2)}
\end{equation}

The average electric polarization only appears in the helicoidal
state and for spins rotating in the $xy$ plane and ${\bf
Q}\parallel {\bf x}$ it is parallel to the $y$ axis:
\begin{equation}
P_y =  \alpha \gamma \chi_e Q \sqrt{\left(T_H - T\right)\left(T_S
+ \Delta/(2\alpha) - T\right)}.
\end{equation}
 Note that since $P_y \propto M_x M_y$ [see
 Eq.(\ref{eq:Phelix})], it has the square root anomaly at
 the ferroelectric transition, even though it is not a
 primary order parameter. Furthermore,  as in proper
 ferroelectrics, the dielectric constant
 $\varepsilon_{yy}$ diverges at $T_H$ and obeys `the
 $1/2$-law' \cite{Landau}:
\begin{equation}
\varepsilon_{yy} \approx \left\{
\begin{array}{cc}
\frac{A}{T - T_H}, & \mbox{for $T > T_H$},\\ \\
\frac{A}{2(T_H-T)}, & \mbox{for $T < T_H$},
\end{array}
\right.
\end{equation}
where $A = 6 \Delta (\gamma \chi_e Q)^2/(\alpha b)$. Although
$P_y$ vanishes in the sinusoidal SDW state, the magnetic
contribution to $\varepsilon_{yy}$ is nonzero up to $T = T_S$:
\begin{equation}
\varepsilon_{yy} = \frac{\left(2 \gamma \chi_e
Q\right)^2}{b}\frac{(T_S - T)}{(T - T_H)},\;\;\;\; \mbox{for
$\,T_H < T < T_S$}.
\end{equation}

\begin{figure}[htbp]
\centering
\includegraphics[width=5.5cm]{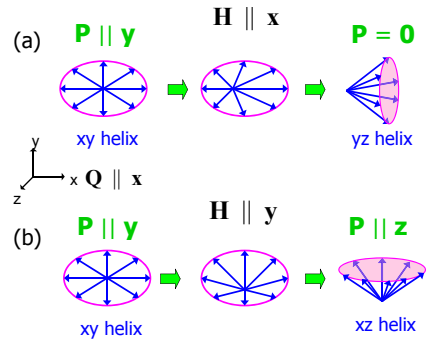}
\caption{Magnetic field behavior of electric polarization for the
model Eq.(\ref{eq:onelayer}). In zero field spins rotate in the
$xy$ plane and ${\bf P} \parallel {\bf y}$. Magnetic field in the
$x$ direction suppresses the polarization, while $H_y$ orients
${\bf P}$ in the $z$ direction. \label{fig:flips}}
\end{figure}

\textit{Behavior in magnetic field:} The salient feature of
ferroelectric magnets, important for technological applications,
is the strong sensitivity of their dielectric properties to
magnetic field, which can suppress electric polarization or change
its direction \cite{Lawes,KimuraLawes}. We first discuss
polarization flops in the model Eq.(\ref{eq:onelayer}). In weak
fields spins rotate in the easy $xy$ plane, so that the spin
rotation axis ${\bf e}_3$ of the helix is parallel to the `hard'
$z$ axis.  For ${\bf Q}
\parallel {\bf x}$ electric polarization is oriented along the $y$
axis. In strong magnetic fields spins form a conical helix with
${\bf e}_3
\parallel {\bf H}$, e.g., $H_x$ will force the spins to rotate in
the $zy$ plane (see Fig.~\ref{fig:flips}a). Such a spin flop will
suppress electric polarization, since for the $zy$-helix ${\bf
e}_3 \parallel {\bf Q}$ and according to Eq.(\ref{eq:Phelix})
${\bf P} = 0$. On the other hand, magnetic field in the $y$
direction favors the rotation of spins in the $xz$ plane, in which
case ${\bf P} \parallel {\bf z}$ (see Fig.~\ref{fig:flips}b).

The magnetic field behavior observed in orthorombic manganites is
somewhat more involved. If we identify the $x$, $y$, and $z$ axes
used in this paper with, respectively, the $b$, $c$, and $a$ axes
of the {\it Pbnm} crystal structure of TbMnO$_3$ ,  then magnetic
field applied in the $x$ and $z$ directions changes the direction
of the electric polarization of TbMnO$_3$ from $y$ to $z$.
According to Eq.(\ref{eq:Phelix}), this corresponds to the change
of the rotation plane from $xy$ to $xz$. It is the flop shown in
Fig.~\ref{fig:flips}b, but induced by magnetic fields with `wrong'
orientations.

\begin{figure}[htbp]
\centering
\includegraphics[width=5cm]{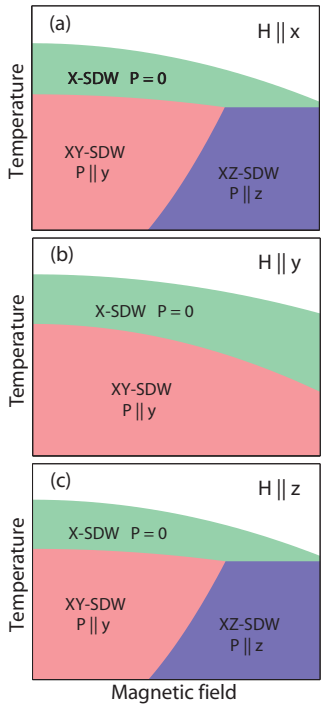}
\caption{Typical phase diagrams of the model
Eq.(\ref{eq:onelayer}) with nonlinear terms
Eq.(\ref{eq:nonlinear}) included for ${\bf H} \parallel {\bf x},
{\bf y}, {\bf z}$ [respectively, panels (a),(b), and (c)].  Here
X-SDW (green) denotes the sinusoidal SDW state with spins along
the $x$ axis, while XY-SDW (pink) and XZ-SDW (blue) denote the
spiral states with spins rotating, respectively, in the $xy$ and
$xz$ planes. \label{fig:phdmf}}
\end{figure}

This unusual behavior is most likely related to the flops of the
strongly anisotropic rare earth  spins, coupled to Mn spins
\cite{KimuraLawes,Kenzelmann}. It can be described
phenomenologically by adding the higher-order anisotropies to
Eq.(\ref{eq:onelayer}), e.g.,
\begin{eqnarray}
\Delta \Phi_m({\bf M}) &=& b_{xy} \left(M_x\right)^2
\left(M_y\right)^2 + b'_{xy} \left(M_x\right)^2 \left(\frac {d
M_y}{dx}\right)^2 \nonumber
\\&+& b_{yz} \left(M_y\right)^2 \left(M_z\right)^2.
\label{eq:nonlinear}
\end{eqnarray}
For positive coefficients the first two terms suppress the
rotation in the $xy$  plane, when magnetic field is applied in the
$x$ or $y$ direction, while the last term suppresses the rotation
in the $yz$ plane. This gives rise to phase diagrams, shown in
Fig.~\ref{fig:phdmf}, which are similar to the ones found for
TbMnO$_3$ \cite{KimuraLawes}. The nonlinear terms also result in a
magnetic field dependence of the helicoidal transition
temperature, which together with the divergency of dielectric
constant at $T_H$ makes $\varepsilon$ strongly field-dependent
(the giant magnetocapacitance effect \cite{Goto}).

\begin{figure}[htbp]
\centering
\includegraphics[width=6cm]{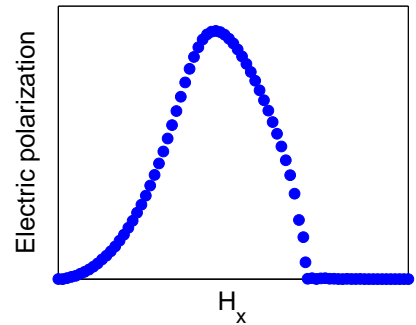}
\caption{Predicted magnetic field dependence of the electric
polarization $|P^y| = |P^z|$ of CaFeO$_3$ for ${\bf H} \parallel
{\bf x}$. \label{fig:CaFeO3}}
\end{figure}

Importantly, magnetic fields required to induce spin flops are of
the order of magnetic anisotropies that can be relatively small
for transition metal ions with filled and half-filled $t_{2g}$
shells, e.g., Cu$^{2+}$, Ni$^{2+}$, Fe$^{3+}$, and Mn$^{3+}$.
Therefore, electric polarization can be flopped by modest magnetic
fields even in spiral magnets with high ordering temperatures,
which may be interesting for applications. Consider, e.g., a {\em
gedanken} experiment on the helimagnet CaFeO$_3$ with $T_N = 115$K
\cite{Kawasaki}. In zero field both the wave vector and the spin
rotation axis of the helix are parallel to the body diagonal,
${\bf Q}, {\bf e}_3
\parallel [ 1, 1, 1]$, and no electric polarization is expected.
Magnetic field of the order of magnetic anisotropies can flop the
orientation ${\bf e}_3$. On the other hand, to change the wave
vector ${\bf Q}$ would require fields of the order of the
antiferromagnetic superexchange between neighboring iron spins
\cite{Mostovoy}, which are much stronger. Thus magnetic field
along one of the crystal axes, e.g. ${\bf H} = H_x \left[ 1, 0,
0\right]$, will induce ${\bf P} = P [0, -1, 1]$. The result of the
numerical calculation of $P(H_x)$, using the Ginzburg-Landau
expansion similar to Eq.(\ref{eq:onelayer}), is shown in
Fig.~\ref{fig:CaFeO3}.

In conclusion, we used simple symmetry arguments to explain
ferroelectric properties and thermodynamics of spiral magnets.
Taking into account the complexity of exchange interactions and
spin orders in these materials, this phenomenological approach
works surprisingly well.

I would like to thank the University of Cologne, where this work
was largely done, for hospitality and Daniel Khomskii for numerous
discussions. The financial support by the DFG (Merkator
fellowship) and MSC$^{\mathit{plus}}$ program is gratefully
acknowledged.

\end{document}